\documentstyle[preprint,aps,version2]{revtex}
\begin{document}

\begin{titlepage} 
\begin{title}
Symmetry aspects of the pion leptoproduction and  the upper limit 
of the Levelt-Mulders asymmetry\thanks{
This work was  supported in part by  the National 
Science Foundation of China.}
\end{title}

\author{Wei Lu$^1$, Jian-Jun Yang$^2$}

\begin{instit}
$^1$ CCAST (World Laboratory), P.O. Box 8730, Beijing 100080, P.R. China 
 and Institute of High Energy Physics, 
 P.O. Box 918(4), Beijing 100039, P.R. China\footnote{Mailing address}, 
(e-mail: luw@bepc3.ihep.ac.cn)
\end{instit}

\begin{instit}
$^2$ Department of Physics, Nanjing Normal University, 
Nanjing 210024, P.R. China  and 
Institute of High Energy Physics, 
 P.O. Box 918(4), Beijing 100039, P.R. China\footnote{Mailing address}, 
(e-mail: yangjj@bepc3.ihep.ac.cn)
\end{instit}

\begin{abstract}
 We  examined the symmetry aspect of 
the semi-inclusive one-pion production  in the 
deep inelastic scattering of a  lepton  beam off 
 an unpolarized nucleon target, 
with an emphasis on the positivity   restrictions on the 
corresponding structure functions. In combination with 
the Callan-Gross-type relation between two twist-two 
structure functions $W_1$ and $W_2$, we  derived an upper 
bound  on the  Levelt-Mulders asymmetry, which occurs 
when the lepton beam is longitudinally polarized. 
\end {abstract}

\end{titlepage} 
\section{Introduction}

At present, one irritating  fact is 
that one has no way to  make precise predictions for the 
properties of hadron-involved processes from the 
first principles, due to lack of reliable handle on 
the QCD soft interactions.  On the other hand, experiences 
tell us that many bulk properties of particle interactions 
are determined by the symmetry of the system, irrelevant of 
the dynamical details. Hence, it is desirable to learn 
as much as possible about the process of interest 
from  its symmetries. In practice, such analyses 
are of important guidance  both for theoretical 
attempts and for  experimental  researches. 

During the past decade, much  theoretical attention 
[1-12] 
has been paid to the 
semi-inclusive pion production in the deep inelastic 
lepton-nucleon scattering.  Moreover, 
some experimental efforts [13,14] have already 
appeared. In studying the deep inelastic scattering  of the 
charged lepton off the nucleon,  one usually 
adopts the so-called one-photon  exchange approximation. 
Accordingly, the  response of the target nucleon to the 
photon probe is  characterized by  a hadronic tensor (see its definition 
in Section II).  One of its symmetry properties is 
reflected in its decomposition 
into Lorentz invariant structure functions.  To our knowledge, 
Mulders \cite{Mulders} was first to   suggest 
the decomposition of the hadronic tensor
for the unpolarized pion leptoproduction. 
In the Mulders parameterization,  there are four independent 
structure functions.   Recently, Levelt and Mulders \cite{lm1} 
worked out a QCD factorization approach to the hadronic tensor,
whose results contain an imaginary part, which cannot 
be accommodated into the early Mulders decomposition. 
At the same time,  Kotzinian  \cite{kotz} 
counted the number of independent 
structure functions in the case of a polarized nucleon target, 
by working in a specific frame and treating the hadronic tensor 
as a  matrix in the Lorentz space. 
According to Ref. \cite{kotz}, there are five independent 
structure functions.  Hence, we feel it necessary to 
clarify the  question how many 
independent structure functions there are 
in the pion leptoproduction with an unpolarized nucleon target.

Furthermore, there is another aspect of the symmetry properties 
of the  pion leptoproduction, i.e., the hadronic tensor 
is subject to some positivity constraints. 
We anticipate that the  positivity  of the hadronic 
tensor for the pion leptoproduction can also
provide  us some useful bounds to the structure functions, 
just like in the case of the deeply inelastic scattering \cite{Doncel}. 
Historically, the positivity  restrictions to 
the pion leptoproduction 
was investigated by Gourdin \cite{Gourdin} as early as in 1972. 
However, the structure functions were parameterized 
in Ref. \cite{Gourdin} in terms of 
the cross sections for the associated  photon-nucleon  reactions 
with the same inclusive-pion final states. 
Since it is now a prevailing  practice  for the 
particle physics community to decompose $directly$ 
the  hadronic  tensors into the Lorentz invariant structure 
functions,  we  recast the positivity  restrictions to the 
pion leptoproduction,   with the hadronic tensor 
parameterized in terms of Lorentz invariant structure 
functions.

In Ref. \cite{lm1}, Levelt and Mulders 
\cite{lm1}  identified a $\langle\sin\phi\rangle$-type
  asymmetry, which comes about 
when  one collides a longitudinally polarized lepton beam on an 
unpolarized nucleon target.  Here $\phi$ is the azimuthal angle of the 
detected pion about the lepton scattering plane. 
As these authors showed, the measurement of such a  single  spin 
asymmetry can allow for a determination of 
some naively-time-reversal-odd  quark fragmentation functions. 
This kind of experiments can be expected to be done 
in the near future  at the high luminosity   facilities, 
such as DESY HERA \cite{HERMES}, CERN LHC \cite{HELP}, and the 
proposed Electron Laboratory for Europe (ELFE) \cite{ELFE}.  Since single 
spin asymmetries are of their own relevance in our understanding 
of the hadron  structure 
and dynamics, it is preferable to learn more about 
the Levelt-Mulders  asymmetry before the relevant experiments come true.

The purpose of this paper is to 
examine  systematically the symmetry aspect of  the 
semi-inclusive one-pion production in the deep 
inelastic scattering of a lepton beam off an unpolarized 
nucleon target.  By applying  Mueller's generalized 
optical theorem,  we show that there are five independent structure 
functions  for the  process considered, which is 
consistent with the counting by Kotzinian. 
Then, we  study  the positivity constraints to these structure functions, 
which are essentially due to the symmetry of the hadronic tensor. 
An important new ingredient  imbedded in  our
positivity analysis  is to combine  the  positivity 
restrictions  with the Callan-Gross relation among the 
structure functions,  which yields an upper bound to
the Levelt-Mulders asymmetry. The upper limits  derived 
eliminate the possibility of observing large Levelt-Mulders 
asymmetries  in certain kinematical regions and serve as 
a judgement for the reliability of the experimental data in the future.

  The  paper is organized as follows.  In Sect. \ref{II} we define 
our kinematics and examine the general symmetry constraints 
to the Lorentz structure  of the hadronic tensor, with an emphasis
on the  fact that time-reversal invariance of electromagnetic 
interactions  does not exert any constraints on the 
hadronic tensor,  owing to the existence of the final-state 
interactions in the inclusively-detected  final state. 
In Sect. \ref{III}, we employ the Mueller theorem to enumerate 
the number of independent structure functions for the
process considered.  In Sect. \ref{IV},  
we  discuss several scenarios for decomposing 
the hadronic tensor.  Sect. \ref{V} is devoted to 
the derivation of  various positivity constraints among the 
structure functions. In Sect. \ref{VI}, we  derive an 
upper limit for the Levelt-Mulders asymmetry by combining 
the positivity constraints and the Callan-Gross relation. 
Sect. {\ref{VII} contains  our concluding remarks.

\section{Hadronic tensor and its general symmetry constraints \label{II} }

The process we will  consider  is the semi-inclusive pion production 
on an unpolarized nucleon target
$$ l(k , s_l ) +N(P) \to l(k^\prime ) +\pi(P_\pi) +X,$$
where the particle momenta are self-explanatory and  $ s_l $ is 
the  spin four-vector of the  incident lepton.  We normalize
the spin vector of the lepton  as $ s_l \cdot  s_l =-1$ for a pure state. 
In the one-photon  exchange approximation, the  differential
cross section can be put into  a Lorentz 
contraction of the  leptonic and hadronic tensors: 
\begin{equation}
\frac{d\sigma(s_l)}{dxdy dz d|{\bf P}_{\pi\perp}|^2d\phi } 
=\frac{\alpha^2 y}{32 M \pi^2 x Q^2 |{\bf P}_{\pi ||}| }
L_{\mu\nu}(k ,  s_l ; k^\prime ) W^{\mu\nu}(q, P, P_\pi), 
\label{cross}
\end{equation}
where  $q=k -k ^\prime$ is the virtual photon momentum, 
 ${\bf P}_{\pi||}$ is  the longitudinal momentum 
of the pion along the direction of the motion of the virtual photon, 
and ${\bf P}_{\pi\perp}$ is  the  corresponding transverse pion  momentum. 
In  our presentation, we will employ the 
virtuality of the probe  photon $Q=\sqrt{-q^2},$ and 
the  energy loss of the lepton in the  target rest frame 
$\nu=P\cdot q/M$ with $M$ the nucleon mass. 
 Moreover, we  adopt the the  scalar variables defined as  
\begin{equation}
x=\frac{-q^2}{2 P\cdot q}, ~ y=\frac{P\cdot q}{P\cdot k }, 
~ z=\frac{P\cdot P_\pi}{P\cdot q}. 
\end{equation}
$|{\bf P}_{\pi||}|$ is related to $|{\bf P}_{\pi\perp}|$ by 
\begin{equation}
|{\bf P}_{\pi||}|= \frac{zQ^2}{2Mx}\sqrt{1-\left(\frac{2Mx}{zQ^2}\right)^2
\Big(
|{\bf P}_{\pi\perp}|^2 +M^2_\pi\Big)}.  
\end{equation} 

In Eq. (\ref{cross}), the leptonic tensor is defined as 
\begin{eqnarray} 
L^{\mu\nu}(k , s_l  ;k^\prime)&\equiv& {\rm Tr} \left[ 
(\rlap/k^\prime  +M_l)\gamma_\mu
 (\rlap/k  +M_l)\gamma_\nu
\frac{1+\gamma_5 \rlap/ s_l }{2}\right]
\nonumber\\ 
&=& -q^2 (-g^{\mu\nu}+\frac{q^\mu q^\nu}{q^2}) 
+4(k ^\mu -\frac{q^\mu}{2})(k ^\nu -\frac{q^\nu}{2})
+2iM_l \epsilon^{\mu\nu\alpha\beta}{q_\alpha}s_{l\beta},
\label{lep}
\end{eqnarray} 
where $M_l$ is the lepton mass. Accordingly, the 
hadronic tensor in this paper is defined as 
\begin{equation} 
W^{\mu\nu}(q,P,P_\pi)=\frac{1}{4\pi}\sum_X 
\int d^4\xi 
\exp(iq\cdot \xi) \langle P| J^\mu (0)
| \pi(P_\pi),X\rangle \langle \pi(P_\pi),X|  J^\nu (\xi)|
P\rangle, 
\end{equation}
where 
the summation over $X$ exhausts all the possible 
final states that contain the chosen pion. 
In our work, the electromagnetic quark current is defined as 
$J^\mu=\sum_f e_f \bar\psi_f \gamma^\mu \psi_f$, with 
$f$ the quark flavor index and $e_f$ the electric charge 
of the  quark  in unit of the electron charge.  
Throughout  we normalize  the one-particle state 
in a relativistic way 
that  $\langle P|P^\prime\rangle =(2\pi)^3 
2E\delta^3 ({\bf P} -{\bf P}^\prime)$. 
Our  conventions are different from those in Refs. 
[4,9,10,12], but there is no  principal difference. 

Because the fundamental interaction vertex is 
electromagnetic in the one-photon exchange approximation, 
the Lorentz structure of the hadronic tensor is 
subject to  all the symmetries that electromagnetic 
interactions observe. Now we examine these symmetry constraints. 

First, the electromagnetic  interaction is gauge invariant, 
which is reflected as the following current conservation conditions: 
\begin{equation} 
q_\mu W^{\mu\nu}(q,P, P_\pi)
=
q_\nu W^{\mu\nu}(q,P,P_\pi)=0. 
\label{gauge}
\end{equation} 

Second, the electromagnetic current is Hermitian,  which leads to 
\begin{equation} 
[W^{\mu\nu}(q,P,P_\pi)]^\ast
=W^{\nu\mu}(q,P,P_\pi). 
\label{hermi}
\end{equation}

Thirdly, the electromagnetic interaction is parity conserved. 
For a generic Lorentz vector $x^\mu$, we 
define $\tilde x^\mu=x_\mu$ following  
Itzykson and Zuber \cite{Zuber}. Then, 
the parity conservation of the electromagnetic 
interaction informs us that 
\begin{equation} 
W^{\mu\nu}(q,P,P_\pi) 
=
W_{\mu\nu}(\tilde q,\tilde P,\tilde P_\pi). 
\label{p}
\end{equation}

Fourthly, the fundamental electromagnetic  vertex is 
invariant under time-reversal transformation. 
We  recall that  time-reversal transformation
includes making a complex conjugation and 
changing the in-state into its corresponding out-state, 
or vice versa. In general, an in-state is related to its 
corresponding out-state by $S$ matrix (operator): 
\begin{equation} 
|\rangle _{\rm in} =S|\rangle _{\rm out}, \label{in-out}
\end{equation}
with $S=1+iT.$  The difference between the in-state 
and its associated out-state is essentially due to the 
final-state interactions described by  $T$, 
the transition matrix (operator). 
Unless the state is composed of an individual particle
or a set of non-interactive particles,  the in-state differs 
from its corresponding out-state.   Hence, 
time-reversal invariance can only tell us 
\begin{equation} 
W^{\mu\nu}(q,P,P_\pi)=
\left[ \frac{1}{4\pi}
\sum \limits_X
\int  d ^4 \xi  e^{i q \cdot \xi}
\langle \tilde P |j^\dagger_\mu (0)
|\pi (\tilde P_\pi),X\rangle _{\rm in ~ in}
\langle \pi(\tilde P_\pi),X|j_\nu(\xi)|
\tilde P \rangle  \right]^\ast. 
\end{equation} 

If one does  not distinguish the in-state from its corresponding 
out-state, then there is the  so-called naive time reversal transformation. 
Under such a simplified time-reversal transformation, there will be 
\begin{equation} 
W^{\mu\nu}(q,P,P_\pi)=
[W_{\mu\nu}(\tilde q,\tilde P,\tilde P_\pi )]^\ast.  \label{t}
\end{equation} 
From Eqs. (\ref{p}) and (\ref{t}), it can be seen that
it is more convenient to  use the adjoint parity-time-reversal 
transformation instead of the individual 
parity and time-reversal transformations. 
For our hadronic tensor, the adjoint parity-time-reversal transformation 
gives rise to 
\begin{equation} 
W^{\mu\nu}(q,P,P_\pi)=
\left[ \frac{1}{4\pi}
\sum \limits_X
\int  d ^4 \xi  e^{iq \cdot \xi}
\langle P|j^{\dagger\mu} (0)
|\pi(P_\pi),X\rangle _{\rm in ~ in}
\langle \pi(P_\pi), X|j^\nu(\xi)|
P\rangle  \right]^\ast. 
\label{pt}
\end{equation} 
Substituting Eq. (\ref{in-out}) into (\ref{pt}),  
we can decompose the hadronic  tensor into  two parts: 
\begin{equation} 
W^{\mu\nu}(q,P,P_\pi)=
W^{(S)\mu\nu}(q,P,P_\pi)
+W^{(A)\mu\nu}(q,P,P_\pi), 
\label{wdec}
\end{equation} 
where $W^{(S)\mu\nu}(q,P,P_\pi)$  survives the naive time-reversal 
transformation but $W^{(A)\mu\nu}(q,P,P_\pi)$  does not. 
The  occurrence of $W^{(A)\mu\nu}(q,P,P_\pi)$  is completely
due to the difference between the in-state  and out-state, 
i.e., the final-state interactions.  By turning off the 
final-state interactions,   one can show  from Eqs. (\ref{hermi}), 
(\ref{p}), and (\ref{t}) that $W^{(S)\mu\nu}(q,P,P_\pi)$ 
is symmetric with respect to indices $\mu$ and $\nu$. 
In principle,  the final-state-interaction-caused contributions
to $W^{\mu\nu}(q,P,P_\pi)$ are $asymmetric$
under the exchange $\mu\leftrightarrow\nu$.  However, 
one can partition   those symmetric contributions 
from the final-state interactions  into $W^{(S)\mu\nu}(q,P,P_\pi)$. 
Therefore, $W^{(A)\mu\nu}(q,P,P_\pi)$ will be antisymmetric 
with respect to $\mu$ and $\nu$, or equivalently, 
odd   under the naive parity-time-reversal transformation. 

At this stage, we  have clarified all the symmetry constraints 
of the hadronic tensor. It seems straightforward to write down 
its general Lorentz decomposition, 
in the complete basis constructed by the Lorentz vectors 
associated with the probe photon, target nucleon, and the 
inclusive pion, along with the metric tensor $g_{\mu\nu}$
and the completely antisymmetric tensor $\epsilon_{\mu\nu\rho\sigma}$.
However, if we set about immediately this task, great risk is 
taken of overcounting or undercounting  the number of 
structure functions as many  terms 
can be  constructed satisfying Eqs. (\ref{gauge}), 
(\ref{hermi}), and (\ref{p}).  Therefore, it is imperative 
to know in advance the number of the 
independent terms  before setting about 
the general  Lorentz expansion of hadronic tensor.

\section{Number of independent structure functions \label{III} }

For counting the number of independent structure functions,
the Mueller theorem \cite{Mueller} supplies us 
with a very convenient method. Let us  truncate the leptonic 
scattering part  of the  pion leptoproduction 
and consider equivalently the inclusive process 
$\gamma^\ast(q,\epsilon) +N(P)\to \pi(P_\pi)+X,$
where $\epsilon^\mu$ is the  polarization vector of the virtual photon 
$\gamma^\ast$.  Obviously, its cross section 
is proportional to $\epsilon^\mu\epsilon^\nu 
W_{\mu\nu}(q,P,P_\pi)$ and can be parameterized in terms 
of a set of independent structure functions. On  the other hand, 
the Mueller  theorem tells us that this 
cross section can be  related to the helicity  amplitudes  for the 
forward three-body scattering 
$
\gamma^\ast +N+ \pi \to \gamma^\ast +N+ \pi \label{three}.
$
Therefore, the  number of the structure functions is equal to that
of the independent forward three-body scattering amplitudes. 
In such a helicity amplitude analysis, the 
unpolarized nucleon can be replaced by a 
spin-zero particle \cite{Gold}.  Then, the helicity amplitude for the 
above forward scattering process is characterized by 
$f_{\lambda_{\gamma^\ast};
\lambda_{\gamma^\ast}^\prime } $,  
where $\lambda_{\gamma^\ast}$ and $\lambda_{\gamma^\ast}^\prime$ 
are the helicities of the virtual photon before and after the scattering. 
Since the virtual photon  has three  helicity states, 
there are $3\times 3 =9$ helicity amplitudes for the 
forward three-body scattering considered.  Obviously, 
not all  of them are independent and they are subject 
to the following parity conservation constraints: 
\begin{equation} 
f_{
-\lambda_{\gamma^\ast};
-\lambda^\prime _{\gamma^\ast}
} =
(-1)^{
\lambda_{\gamma^\ast}
-\lambda_{\gamma^\ast}^\prime
}
f_{
\lambda_{\gamma^\ast}; 
\lambda^\prime _{\gamma^\ast}
}. 
\end{equation} 
Hence there  are  only  five  independent  helicity amplitudes. 
Correspondingly, there are  five  structure functions 
in the decomposition of the hadronic tensor. 
Although time reversal invariance  does not 
yield any further constraints,  we can still learn 
some useful information about the naive-parity-time-reversal 
properties of structure functions.  If there were no 
final-state interactions,  there would be the 
relation like 
$f_{\lambda_{\gamma^\ast}; 
\lambda_{\gamma^\ast}^\prime}
=
f_{\lambda_{\gamma^\ast}^\prime;
\lambda_{\gamma^\ast}},$   which leads to one more 
restriction among the  five independent helicity amplitudes.
Therefore,  we conclude that $W^{(S) \mu\nu}(q,P,P_\pi)$  
and  $W^{(A) \mu\nu}(q,P,P_\pi)$  contain four  and one structure 
functions, respectively.

\section{Several Lorentz decompositions of the hadronic tensor \label{IV}}

Nowadays, it is a common practice  for the particle physics community
to decompose the hadronic tensor into Lorentz 
 invariant structure functions.  On the basis of the discussion 
in the last two sections, one can easily construct the following 
most general Lorentz decomposition for  our hadronic tensor: 
\begin{eqnarray}
W^{\mu\nu} &(& q,P,P_\pi)
=\frac{1}{P\cdot q} (-g^{\mu\nu}+\frac{q^\mu q^\nu}{q^2})w_1
+\frac{1}{q^2(P\cdot q)} 
(P^\mu - \frac{P\cdot q}{q^2}q^\mu) 
(P^\nu - \frac{P\cdot q}{q^2}q^\nu)w_2
\nonumber  \\
& & + \frac{1}{q^2(P\cdot q)} \left[
(P^\mu - \frac{P\cdot q}{q^2}q^\mu) 
(P^\nu_\pi  - \frac{P_\pi\cdot q}{q^2}q^\nu) 
+(P^\nu - \frac{P\cdot q}{q^2}q^\nu) 
(P^\mu_\pi  - \frac{P_\pi\cdot q}{q^2}q^\mu) \right ]w_3 
\nonumber  \\
& & + \frac{1}{q^2(P\cdot q)} 
(P^\mu_\pi - \frac{P_\pi\cdot q}{q^2}q^\mu) 
(P^\nu_\pi  - \frac{P_\pi\cdot q}{q^2}q^\nu) w_4
\nonumber  \\
& & + \frac{1}{q^2(P\cdot q)} \left[
(P^\mu - \frac{P\cdot q}{q^2}q^\mu) 
(P^\nu_\pi  - \frac{P_\pi\cdot q}{q^2}q^\nu) 
-(P^\mu - \frac{P\cdot q}{q^2}q^\mu) 
(P^\nu_\pi  - \frac{P_\pi\cdot q}{q^2}q^\nu) \right]\hat w
\label{de}
\end{eqnarray}
where  $w_1$, $w_2$, $w_3$, $w_4$ and $\hat w$ are 
dimensionless structure functions, dependent on $q^2$, 
$P\cdot q$, and $P\cdot P_\pi$. This decomposition 
is irrelevant of the frame in which one works. 
Here we note that the hadronic tensor for inclusive  one-particle 
leptoproduction  
has an energy dimension two lower than 
its deep inelastic scattering counterpart.

 In practice,  one usually chooses a specific frame 
in which to work.  If one lets the $\hat z$-axis 
be along the direction of  the motion of the  probe 
photon and puts the $\hat x$-axis in the  lepton scattering plane, 
one can build another decomposition of the hadronic tensor. 
In this case, one can introduce an auxiliary four-momentum, 
\begin{equation} 
P^\mu_{\pi\perp}= (0, |{\bf P}_{\pi\perp}|\cos \phi, 
|{\bf P}_{\pi\perp}|\sin \phi, 0), 
\end{equation}
where ${\bf P}_{\pi\perp}$ is the pion  transverse momentum
with respect to the travelling direction of 
 virtual photon and $\phi$ is 
the azimuthal angle of the detected pion.  Obviously, 
$P^\mu_{\pi\perp}$ satisfies $P_{\pi\perp}\cdot q=0$. 
By substituting $P_{\pi\perp}$  for 
$(P_\pi  - \frac{P_\pi\cdot q}{q^2}q)$ in Eq. (\ref{de}), 
one arrives at the following decomposition: 
\begin{eqnarray}
W^{\mu\nu}(q,P,P_\pi)&
=&\frac{1}{P\cdot q} (-g^{\mu\nu}+\frac{q^\mu q^\nu}{q^2})W_1
+\frac{1}{q^2(P\cdot q)} 
(P^\mu - \frac{P\cdot q}{q^2}q^\mu) 
(P^\nu - \frac{P\cdot q}{q^2}q^\nu)W_2
\nonumber  \\
& & + \frac{1}{q^2(P\cdot q)} \left[
(P^\mu - \frac{P\cdot q}{q^2}q^\mu) P^\nu_{\pi\perp}
+P^\mu_{\pi\perp}(P^\nu - \frac{P\cdot q}{q^2}q^\nu)\right]W_3
+
\frac{1}{q^2(P\cdot q)} 
P^\mu_{\pi\perp} P^\nu_{\pi\perp}W_4 
\nonumber  \\
& & +\frac{i}{q^2(P\cdot q)}  \left[
(P^\mu - \frac{P\cdot q}{q^2}q^\mu) P^\nu_{\pi\perp}
-P^\mu_{\pi\perp}(P^\nu - \frac{P\cdot q}{q^2}q^\nu)\right]
\hat W\label{dec}
\end{eqnarray}
where  $W_1$, $W_2$, $W_3$, $W_4$ 
and $\hat W$ are  dimensionless structure functions, 
dependent on $q^2$, $P\cdot q$, and
 ${\bf P}^2_{\pi\perp}$. The advantage of this 
decomposition is that the dependence of cross section 
on the transverse momentum of the detected pion 
can be easily worked out in analytical calculations.
Nevertheless, one has to be aware that this decomposition 
is frame-dependent. 

The  QCD  factorization results in  Ref. \cite{lm1}
can be  tailored into  the above  decomposition, Eq. (\ref{dec}). 
In the literature,  it is Mulders \cite{Mulders} who first worked out 
the terms associated with $W_1$, $W_2$, $W_3$, $W_4$. 
Indeed, the $\hat W$ term, because of its antisymmetric property,  
does not make contributions to cross section 
when the incident lepton beam  is unpolarized, 
which is the very case discussed in Ref. \cite{Mulders,lm2}. 
As has been explained in Sects. \ref{II} and \ref{III}, however, 
the term associated with $\hat W$ incorporates the final-state 
interactions in the inclusively detected state,  so its 
existence does not depend upon whether the initial-state 
beam is polarized or not.  Hence, we claim that it was inappropriate 
in Refs. \cite{Mulders,lm2} to  ignore the final-state 
effects without precautions.

Since the hadronic tensor is a $4\times 4$ matrix 
in the Lorentz  space, one can also parameterize  it 
in terms of its specific matrix elements. Such an 
analysis has  already been  done by Kotzinian \cite{kotz}, 
who discussed the more complicated case with a polarized 
nucleon target.  However,  the Lorentz 
invariance  of the structure functions in 
such  parameterizations is not manifest. For 
comparison,  we note that  five  spin-independent 
structure functions, under distinct disguises, 
were also identified in Ref. \cite{kotz}.  Among them, 
the imaginary part of a matrix element, Im$ H^{(0)}_{01}$, 
corresponds to $\hat w$ in (\ref{de}) and 
$\hat W$ in (\ref{dec}).

In fact, one can also construct other Lorentz 
decompositions. Because adopting different conventions, 
different authors usually have different decompositions. 
However, the  number of independent structure functions 
should always be  fixed  because it is  a reflection of the 
symmetries of the hadronic tensor.  
In principle, one can  establish the connections among 
his own structure functions, on the one hand, and 
those by other authors, on the other hand.

\section{Positivity constraints to the  hadronic tensor \label{V}} 

In the rest of this paper, we work in the target rest
frame, with the direction of the motion of the photon probe
in the $\hat z$-axis  and  the lepton scattering plane 
in the $\hat x-\hat z$ plane.  Correspondingly, we 
adopt Eq. (\ref{dec}) as our decomposition of the hadronic tensor.

 The starting point for our positivity analysis is 
the Hermiticity of the electromagnetic  current, 
$J^{\mu\dagger}=J^\mu$. For an arbitrary  Lorentz vector $a^\mu$, 
one can show  
\begin{equation} 
W_{\mu\nu}(q,P,P_\pi) a^{\ast \mu} a^\nu 
\propto \sum\limits_X \delta^4 (q-P-P_\pi-P_X)
|\langle \pi(P_\pi),X|a\cdot J|P\rangle |^2, 
\end{equation} 
so $W^{\mu\nu}(q,P,P_\pi)$ is a semi-positive definite form: 
\begin{equation} 
W_{\mu\nu}(q,P,P_\pi) a^{\ast \mu} a^\nu \geq 0. \label{poi}
\end{equation} 
As a consequence, the relevant 
structure  functions are constrained by some   positivity conditions.

For a  generic Lorentz vector,  one can always expand it over 
a complete set of  bases constructed by four other 
$independent$  vectors.  Of course,  one can choose  one  vector of the
bases to be  proportional to the momentum of the virtual photon 
and the other three as the three polarization vectors of the probe photon: 
\begin{equation} 
e^\mu_1= -\frac{1}{\sqrt 2} (0, 1, +i, 0),
\end{equation}  
\begin{equation} 
 e^\mu_2= +\frac{1}{\sqrt 2} (0, 1, -i, 0),
\end{equation}  
\begin{equation} 
 e^\mu_3= \frac{1}{Q}(\sqrt{\nu^2+Q^2}, 0, 0, \nu). 
\end{equation} 
Notice that these three polarization vectors are 
orthonormal, namely, 
\begin{equation} 
e^\ast_1 \cdot e_1 =e^\ast_2 \cdot e_2 =-e^\ast_3 \cdot e_3 =-1,
\end{equation}  
\begin{equation} 
e^\ast_i \cdot e_j =0,~{\rm with}~i\not= j.
\end{equation} 
In  addition, they satisfy the Lorentz condition 
\begin{equation} 
e_i\cdot q=0,  ~~i=1, 2, 3. 
\end{equation}

Obviously,  by letting $a^\mu=q^\mu$  one can  gain 
only an identity $0\equiv 0$, which  reflects the 
current conservation of the electromagnetic interaction. 
Taking either  $a^\mu=e^\mu_1$  or $a^\mu=e^\mu_2$, 
however, one can obtain the following restrictions: 
\begin{equation} 
M\nu e^{\ast\mu}_1 e^\nu_1 W_{\mu\nu}
=
M\nu e^{\ast\mu}_2 e^\nu_2 W_{\mu\nu}
=
W_1 - \frac{|{\bf P}_{\pi\perp}|^2}{2Q^2} W_4
\geq 0,\label{9}
\end{equation}
On the other hand, one has  with $a^\mu=e^\mu_3$
\begin{equation} 
M\nu e^{\ast\mu }_3 e^\nu_3 W_{\mu\nu}
=-W_1 -\frac{M^2(\nu^2+Q^2)}{Q^4} W_2 
\geq 0.  \label{10}
\end{equation} 
However, both (\ref{9}) and (\ref{10}) are only the  direct 
consequences of the positivity of $W^{\mu\nu}(q,P,P_\pi)$. 
In other words, they are only necessary conditions. 

 As a matter of fact, the sufficient and necessary conditions 
for the positivity of a matrix  are that all  of its submatrices 
have  semi-positively finite determinants \cite{soffer}. Note that
the hadronic tensor considered is a matrix in the Lorentz space,
\begin{equation} 
W^{\mu\nu}(q,P,P_\pi)= 
\left(
\begin{array}{cccc}
W^{00} & W^{01} &W^{02} & W^{03} \\ 
W^{10} & W^{11} &W^{12} & W^{13} \\ 
W^{20} & W^{21} &W^{22} & W^{23} \\ 
W^{30} & W^{31} &W^{32} & W^{33} 
\end{array}
\right).
\end{equation} 
In order to investigate the positivity  restrictions on the  pion 
leptoproduction,  we write out explicitly the elements of $W^{\mu\nu}
(q,P,P_\pi)$ in our coordinate system: 
\begin{eqnarray} 
W^{00}&=&-\frac{\nu^2+Q^2}{M\nu Q^2}W_1 
-\frac{M(\nu^2+Q^2)^2}{\nu Q^6}W_2,\\ 
W^{11}&=&\frac{W_1}{M\nu} 
 -\frac{ |{\bf P}_{\pi\perp}|^2}{M\nu Q^2 }\cos^2\phi W_4,\\
W^{22}&=&\frac{W_1}{M\nu} 
-\frac{ |{\bf P}_{\pi\perp}|^2}{M\nu Q^2 }\sin^2\phi W_4,\\
W^{33}&=&-\frac{\nu}{MQ^2}W_1-
\frac{M\nu(\nu^2+Q^2)}{Q^6} W_2,\\
W^{01}&=&W^{\ast 10}=-
\frac{(\nu^2+Q^2) |{\bf P}_{\pi\perp}|}{\nu Q^4}\cos\phi (W_3 +i\hat W),\\
W^{02}&=&W^{\ast 20}=-\frac{(\nu^2+Q^2) |{\bf P}_{\pi\perp}|}
{\nu Q^4}\sin\phi 
(W_3 +i\hat W),\\
W^{03}&=&W^{30}= -\frac{\sqrt{\nu^2 +Q^2}}{MQ^2}W_1 
-\frac{M\sqrt{(\nu^2+Q^2)^3}}{Q^6}W_2,\\
W^{12}&=&W^{21}= -\frac{ |{\bf P}_{\pi\perp}|^2}{M\nu Q^2}
\cos\phi\sin\phi W_4,\\
W^{31}&=&W^{\ast 13}
=-\frac{\sqrt{\nu^2+Q^2} |{\bf P}_{\pi\perp}|}{Q^4}\cos\phi (W_3-i \hat W),\\
W^{32}&=&W^{\ast 23}=-\frac{\sqrt{\nu^2+Q^2} 
|{\bf P}_{\pi\perp}|}{Q^4}\sin\phi (W_3-i \hat W). 
\end{eqnarray} 
Now we  are in the position to 
examine the necessary and sufficient 
positivity  conditions  for our hadronic tensor. 

 First, the  determinant of $W^{\mu\nu}(q,P,P_\pi)$ itself must be
semi-definitely positive. However, our explicit calculation shows that  
${\rm Det}[ W^{\mu\nu}(q,P,P_\pi)]=0$. 
This occurs by no means accidentally 
for it  reflects the  electromagnetic  gauge 
invariance of the hadronic tensor.  Because $q_\mu W^{\mu\nu}(q,P,P_\pi)
=q_\nu W^{\mu\nu}(q,P,P_\pi)=0$, $W^{\mu\nu}(q,P,P_\pi)$ is at most 
at rank three.  Correspondingly, the determinant  of 
$W^{\mu\nu}(q,P,P_\pi)$ vanishes identically.

Second, two $3\times 3$ submatrices of $W^{\mu\nu}(q,P,P_\pi)$ 
must be semi-definitely positive, i.e.,
\begin{equation} 
\left|
\begin{array}{ccc}
W^{00} & W^{01}& W^{02}\\ 
W^{10} & W^{11}& W^{12}\\ 
W^{20} & W^{21} &W^{22} 
\end{array}
\right|\geq 0, 
\end{equation} 
\begin{equation} 
\left|
\begin{array}{ccc}
W^{11}& W^{12} &W^{13}\\ 
W^{21} &W^{22} & W^{23}\\
W^{31} &W^{32} & W^{33} 
\end{array}
\right|\geq 0.
\end{equation} 
Herewith we obtain  a restriction among five structure functions 
\begin{equation}
-W_1 \left[
W_1 +\frac{M^2(\nu^2+Q^2)}{Q^4} W_2 
\right] 
\left[
W_1 -\frac{|{\bf P}_{\pi\perp}|^2}{Q^2}W_4
\right]
-\frac{M^2(\nu^2+Q^2)
|{\bf P}_{\pi\perp}|^2}{Q^6}W_1 (W_3^2+\hat W^2) \geq 0.
\label{toad}
\end{equation}

  Thirdly, the determinants of three $2\times 2$  submatrices are 
semi-definitely positive, i.e., 
\begin{equation} 
\left|
\begin{array}{cc}
W^{00} & W^{01}\\ 
W^{10} & W^{11} 
\end{array}
\right|\geq 0, 
\end{equation} 
\begin{equation} 
\left|
\begin{array}{cc}
W^{11} & W^{12}\\ 
W^{21} & W^{22} 
\end{array}
\right|\geq 0, 
\end{equation} 
\begin{equation} 
\left|
\begin{array}{cc}
W^{22} & W^{23}\\ 
W^{32} & W^{33}\label{x3}
\end{array}
\right|\geq 0.
\end{equation} 
 In our parameterization, 
these three inequalities assume the following forms: 
\begin{eqnarray} 
~~~~~~& &-W_1\left[
W_1 +\frac{M^2(\nu^2+Q^2)}{ Q^4} W_2 
\right] 
\left[
W_1 -\frac{ |{\bf P}_{\pi\perp}|^2}{Q^2}\cos^2\phi W_4
\right] \nonumber\\
~~~~~~& &~~~~~~~
-
\frac{M^2(\nu^2+Q^2)}{Q^6} |{\bf P}_{\pi\perp}|^2
 \cos^2\phi W_1  (W_3^2+\hat W^2) \geq 0,
\label{l1}
\end{eqnarray} 
\begin{equation}
W_1 \left(
W_1 -\frac{|{\bf P}_{\pi\perp}|^2}{Q^2}W_4 \right) \geq 0,
\label{l2}
\end{equation}
\begin{eqnarray}
~~~~~~~& &-W_1\left[
W_1 +\frac{M^2(\nu^2+Q^2)}{Q^4} W_2 
\right] 
\left[
W_1 -\frac{ |{\bf P}_{\pi\perp}|^2}{Q^2}\sin^2\phi W_4
\right] \nonumber\\
~~~~~~~& &~~~~~~
-
\frac{M^2(\nu^2+Q^2)}{Q^6} |{\bf P}_{\pi\perp}|^2
 \sin^2\phi W_1(W_3^2+\hat W^2) \geq 0.
\label{l3}
\end{eqnarray} 
Simply letting $\cos\phi=1$ in (\ref{l1}) or $\sin\phi=1$ in  (\ref{l3}),
two corresponding inequalities  reduce to (\ref{toad}). However, 
combining (\ref{l1}) with (\ref{l3}) will give rise to 

\begin{equation}
-W_1 \left[
W_1 +\frac{M^2(\nu^2+Q^2)}{Q^4} W_2 
\right] 
\left[
W_1 -\frac{|{\bf P}_{\pi\perp}|^2}{2 Q^2}W_4
\right]
-\frac{M^2(\nu^2+Q^2)
 |{\bf P}_{\pi\perp}|^2}{2 Q^6}W_1 (W_3^2+\hat W^2) \geq 0.
\label{td1}
\end{equation}
If taking $\cos\phi=0$ in (\ref{l1}) or $\sin\phi=0$ in  (\ref{l3}),
one will  regain (\ref{10}). 

  Last, each of the diagonal elements  of $W^{\mu\nu}(q,P,P_\pi)$ 
has to be  semi-definitely positive, i.e., 
\begin{equation} 
W^{00} \geq 0,~ 
W^{11} \geq 0,   ~
W^{22} \geq 0,    ~
W^{33} \geq 0.
\end{equation} 
Accordingly, we  obtain  (\ref{10}) as well as the following  
two inequalities: 
\begin{eqnarray} 
W_1 -\frac{ |{\bf P}_{\pi\perp}|^2}{Q^2}\cos^2\phi W_4&\geq & 0, \label{a}\\
W_1-\frac{ |{\bf P}_{\pi\perp}|^2}{Q^2}\sin^2\phi W_4 & \geq &0.  \label{b}
\end{eqnarray} 
 Either by  letting   $\cos\phi=0$  in (\ref{a}) 
 or $\sin\phi=0$ in (\ref{b}), one has  
\begin{equation}
W_1\geq 0. \label{sign}
\end{equation} 
At this stage, we observe that (\ref{toad}) through (\ref{td1}) can 
be divided  safely by $W_1$ without changing 
the direction of the  inequality sign. 
Furthermore, (\ref{10}) in combination with (\ref{sign}) implies that 
\begin{equation}
W_2\leq 0.
\end{equation} 
As one sets $\cos \phi =1$  in (\ref{a}) 
or $\sin \phi=1$  in (\ref{b}), it will  lead to 
\begin{equation}
W_1 -\frac{ |{\bf P}_{\pi\perp}|^2}{Q^2} W_4\geq  0. \label{ax}
\end{equation} 
When adding (\ref{a}) to  (\ref{b}), however, one recovers (\ref{9}).

\section{Upper limits of the Levelt-Mulders asymmetry  \label{VI}}

   Now  we discuss  the  phenomenological implications of the 
above positivity constraints.  In principle, structure functions 
$W_i$ ($i=1,\cdots, 4)$ can be measured  with an unpolarized 
lepton beam while the measurement of  $\hat  W$ requires 
the  polarization  of the incident beam. 
As Levelt and Mulders  have clarified \cite{lm2},  
the determination of $\hat W$ can be done by measuring 
a $\langle\sin \phi\rangle$ asymmetry 
of the considered process in  the case that the lepton beam 
is polarized longitudinally. 

Substituting Eqs.  (\ref{lep}) and (\ref{dec})  into (\ref{cross})
and completing the Lorentz contractions, one has 
\begin{eqnarray} 
\frac{d\sigma(s_l)}
{dxdydz d|{\bf P}_{\pi\perp}|^2d\phi } 
&=&\frac{\alpha^2 y}{32 M \pi^2 x Q^2 {\bf P}_{\pi ||} }
\Big[4x W_1 -\frac{2\kappa^2}{x y^2} 
W_2+\frac{ 4\kappa (2-y)}{\eta y^2}\frac{|{\bf P}_{\pi\perp}|}{Q} 
 \cos\phi  W_3 
\nonumber \\  
& &  -2 x \left( 1+\frac{4\kappa^2}{\eta^2 y^2}
\cos^2\phi \right) \left(
\frac{|{\bf P}_{\pi\perp}|}{Q}\right)^2 W_4 
-8xM M_l\frac{
 {\bf s}_l\cdot {\bf q} \times {\bf P}_{\pi\perp}
}{Q^4} 
 \hat W \Big],
\label{cs}
\end{eqnarray} 
where 
\begin{equation}
\kappa=\sqrt{1-y-\displaystyle 
\frac{M^2x^2y^2}{Q^2}}, \quad 
\eta=\sqrt{1+\displaystyle\frac{4M^2x^2}{Q^2}}.
\end{equation} 

 Now we define a single  spin asymmetry as 
\begin{equation} 
A\equiv \frac{
\displaystyle
\frac{d\sigma(s_l)}
{dxdydz d|{\bf P}_{\pi\perp}|^2d\phi } 
-
\displaystyle
\frac{d\sigma(-s_l)}
{dxdydz d|{\bf P}_{\pi\perp}|^2d\phi } 
}
{
\displaystyle
\frac{d\sigma(s_l)}
{dxdydz d|{\bf P}_{\pi\perp}|^2d\phi } 
+
\displaystyle
\frac{d\sigma(-s_l)}
{dxdydz d|{\bf P}_{\pi\perp}|^2d\phi } 
}. \label{defi} 
\end{equation} 
Substituting Eq. (\ref{cs}) into (\ref{defi}), one has 
\begin{equation} 
A= \frac{
-8xM M_l
 {\bf s}_l\cdot {\bf q} \times {\bf P}_{\pi\perp} \hat W} 
{Q^4\left[
4x W_1 -
\displaystyle
\frac{2\kappa^2}{x y^2} 
W_2 
+
\displaystyle
\frac{ 4\kappa (2-y)}{\eta y^2}
\displaystyle
\frac{|{\bf P}_{\pi\perp}|}{Q} 
 \cos\phi  W_3  -2 x 
\left( 1+\displaystyle
\frac{4\kappa^2}{\eta^2 y^2}
\cos^2\phi \right) \left(
\displaystyle
\frac{|{\bf P}_{\pi\perp}|}{Q}\right)^2 W_4 \right]
}.
\end{equation} 
In the case  of longitudinal polarization, 
the spin four-vector  of the beam  lepton 
is related to its momentum  via 
\begin{equation}
\lim _{M_l\to 0}
~M_l  s_l ^\mu = 2\lambda_l k ^\mu, \label{relat}
\end{equation} 
with $\lambda_l$ being the lepton helicity.  Correspondingly, there is 
the following Levelt-Mulders asymmetry 
\begin{equation} 
A_L\equiv \frac{
\displaystyle
\frac{d\sigma(\lambda_l= +\frac{1}{2})}
{dxdydz d|{\bf P}_{\pi\perp}|^2d\phi } 
-
\displaystyle
\frac{d\sigma(\lambda_l= -\frac{1}{2})}
{dxdydz d|{\bf P}_{\pi\perp}|^2d\phi } 
}
{
\displaystyle
\frac{d\sigma(\lambda_l= +\frac{1}{2})}
{dxdydz d|{\bf P}_{\pi\perp}|^2d\phi } 
+
\displaystyle
\frac{d\sigma(\lambda_l= -\frac{1}{2})}
{dxdydz d|{\bf P}_{\pi\perp}|^2d\phi } 
}. \label{defimul} 
\end{equation}
After a little algebra, we have 
\begin{equation}
A_L= \frac{
-\displaystyle
\frac{4\kappa }{y}
\displaystyle
\frac{|{\bf P}_{\pi\perp}|}{Q} \sin\phi
\hat W} 
{
4x W_1 -
\displaystyle
\frac{2\kappa^2}{x y^2} 
W_2 
+
\displaystyle
\frac{ 4\kappa (2-y)}{\eta y^2}
\displaystyle
\frac{|{\bf P}_{\pi\perp}|}{Q} 
 \cos\phi  W_3  -2 x 
\left( 1+\displaystyle
\frac{4\kappa^2}{\eta^2 y^2}
\cos^2\phi \right) \left(
\displaystyle
\frac{|{\bf P}_{\pi\perp}|}{Q}\right)^2 W_4
}.
\end{equation} 
Because $A_L$ 
is essentially proportional to $\langle {\bf s}_l\cdot 
{\bf q}\times {\bf P}_{\pi\perp}\rangle$,  it has 
the largest values at $\sin\phi=1$, i.e, when the pion momentum 
has no   transverse components in the lepton scattering plane.

 Now we  bound $A_L$ by  use  of the results derived from the 
positivity analysis. From (\ref{toad}) and (\ref{sign}),  
we have  by letting $W_3=0$
\begin{equation}
\frac{|{\bf P}_{\pi\perp}|}{Q}|\hat W|\leq 
\sqrt{ -\left[
\frac{4x^2Q^2}{Q^2+4x^2M^2}
W_1 + W_2 
\right] 
\left[
W_1 -\frac{ |{\bf P}_{\pi\perp}|^2}{Q^2} W_4
\right]}. \label{wolf}
\end{equation}
Correspondingly,  there will be 
\begin{equation}
|A_L|\leq  \frac{
\displaystyle
\frac{4\kappa }{y}
\displaystyle
\sqrt{ -\left[
\frac{4x^2Q^2}{Q^2+4x^2M^2}
W_1 + W_2 
\right] 
\left[
W_1 -\frac{ |{\bf P}_{\pi\perp}|^2}{Q^2}W_4
\right]} \sin\phi  
} 
{\left|
4x W_1 -
\displaystyle
\frac{2\kappa^2}{x y^2} 
W_2 
+
\displaystyle
\frac{ 4\kappa (2-y)}{\eta y^2}
\displaystyle
\frac{|{\bf P}_{\pi\perp}|}{Q} 
 \cos\phi  W_3  -2 x 
\left( 1+\displaystyle
\frac{4\kappa^2}{\eta^2 y^2}
\cos^2\phi \right) \left(
\displaystyle \frac{|{\bf P}_{\pi\perp}|}{Q}\right)^2 W_4 \right|
}. \label{resl}
\end{equation} 

As a rough approximation, we may  neglect all 
the $Q$-power suppressed  effects as compared 
to $O(Q^0)$ quantities. More concretely, we 
drop out all $W_3$- and $W_4$-terms and 
substitute $\sqrt{1-y}$ for $\kappa$ in (\ref{resl}).
As a result, 
\begin{equation}
|A_L|\leq  \frac{
2x y \sqrt{1-y}
\displaystyle
\sqrt{ -W_1\left[
\frac{4x^2Q^2}{Q^2+4x^2M^2}
W_1 + W_2 
\right] 
} \sin\phi 
} 
{|2x^2 y^2 W_1 -(1-y) W_2| }. \label{resll}
\end{equation} 

At this stage, we employ the Callan-Gross-type relation between 
$W_1$ and $W_2$ 
\begin{equation} 
W_1 +\frac{1}{4x^2}W_2=0  \label{Callan}
\end{equation} 
 to simplify  further the right-hand side of (\ref{resll}). 
The verification of Eq. (\ref{Callan})  can  be very easily done 
in the naive quark-parton model without intrinsic transverse 
parton momentum. Actually, it can also be  obtained   simply from  the
fact that the virtual photon  probe tends to be 
transversely polarized  in the high energy limit. Put it in another way, 
\begin{equation} 
e^{\ast\mu}_3 e^\nu_3 W_{\mu\nu}(q, P,P_\pi)
\to 0~{\rm as ~}Q\to\infty~{\rm with}~ x ~{\rm fixed}.
\end{equation} 

Inserting Eq. (\ref{Callan}) into (\ref{resll}),  we arrive at 
the following upper limit for the considered asymmetry
\begin{equation}
|A_L|\leq  \frac{
4xy \sqrt{1-y} \displaystyle \frac{M}{Q} \sin\phi 
} {(y-1)^2 +1 }. \label{final} 
\end{equation} 
Since both $W_3$ and $\hat W$ contribute at one-power 
suppressed level, i.e., at twist three, 
it should be stressed that (\ref{final}) is an amplified upper bound 
for $A_L$. The reason is that in deriving (\ref{wolf}) 
from (\ref{toad}) and (\ref{sign}),  we have assumed $W_3=0$. 
Because experiments on spin asymmetries are usually subject 
to  large  statical errors, this positivity constraint
can be taken as a very useful guide to judge the reliability
of the experimental results. 

On the other hand, (\ref{final}) simply 
informs  us  that the chance to measure 
the Levelt-Mulders asymmetry  is very faint
 in some kinematical domain. To be illustrative, we draw in Fig. 1
the derived upper limit of the Levelt-Mulders asymmetry $A_L$
versus  the  fraction of the lepton energy loss
$y$, with  $x=0.5$, $\sin\phi=1$ and $Q=5M$
at which  the perturbative QCD can be applicable so that
the Callan-Gross relation  is reliable.  

\section{Concluding remarks \label{VII}} 

Obviously, our discussion can be generalized to 
the case of the lepton beam being transversely polarized. 
However, Eq. (\ref{relat}) will no longer hold  and 
the  corresponding spin asymmetry  will be  $M_l/Q$-suppressed. 
 We do  not 
carry out such an  extension  for it  will not supply 
us with any practical experimental guidance. 
In principle, our discussion can also 
 be  generalized  to the case 
in which the nucleon is polarized 
and even to the case  the  semi-inclusively detected hadron 
is a baryon, to say,  a $\Lambda$ hyperon with its spin state 
monitored.  However, such a generalization  will be of less 
relevance because a  large number of 
structure  functions will be involved in  decomposing  the 
corresponding hadronic tensors.

 In  summary, we  have examined  systematically the 
symmetry properties of the semi-inclusive 
one-pion production  induced by a charged lepton beam 
on an unpolarized nucleon target, 
with an emphasis on the  positivity constraints 
to the structure functions. 
We found that due to  the positivity of the hadronic 
tensor, the signs of two twist-two structure functions 
$W_1$ and $W_2$
can be determined. Moreover, there exists 
an inequality restricting five structure functions. 
This restriction, in connection to  the Callan-Gross relation 
between $W_1$ and $W_2$,  yields an  upper bound on 
the Levelt-Mulders asymmetry.

{\it Acknowledgement} One of the authors (W.L)
would like to thank  A. Kotzinian, P. J. Mulders,  and 
 J. Soffer for useful correspondence.

\centerline{\bf \huge Figure Caption} 

Figure 1.  The upper bound  of the Levelt-Mulders asymmetry 
($A_L$)  versus the fraction of the lepton energy loss  ($y$). 
The Bjorken $x$ is taken  to be 0.5,  the pion azimuthal angle 
($\phi$) to be 90$^\circ$ with respect to the lepton scattering plane, 
and the momentum transfer ($Q$) to be five times the nucleon mass ($M$). 
\end{document}